\documentclass[prc,twocolumn,showpacs,amsmath,amssymb,superscriptaddress,floatfix,nofootinbib,10pt]{revtex4-1}

\usepackage{slashed}
\usepackage{mathrsfs,bm}
\usepackage{txfonts}
\usepackage{amssymb}
\usepackage{indentfirst}
\usepackage{graphicx,booktabs}
\usepackage{multirow}
\usepackage{overpic}
\usepackage{color}
\usepackage{amssymb}
\usepackage[utf8]{inputenc}

\begin{document}
\title{One pion exchange and the quantum numbers of the $P_c(4440)$ and $P_c(4457)$ pentaquarks}

\author{Manuel Pavon Valderrama}\email{mpavon@buaa.edu.cn}
\affiliation{School of Physics and Nuclear Energy Engineering, \\
Beihang University, Beijing 100191, China} 
\affiliation{
International Research Center for Nuclei and Particles in the Cosmos and \\
Beijing Key Laboratory of Advanced Nuclear Materials and Physics, \\
Beihang University, Beijing 100191, China} 

\date{\today}
\begin{abstract}
  The LHCb collaboration has recently discovered three pentaquark-like states
  --- the $P_c(4312)$, $P_c(4440)$ and $P_c(4457)$ --- close to
  the $\bar{D} \Sigma_c$ and the $\bar{D}^* \Sigma_c$
  meson-baryon thresholds.
  The standard interpretation is that they are heavy antimeson-baryon molecules.
  Their quantum numbers have not been determined yet, which implies
  two possibilities for the $P_c(4440)$ and $P_c(4457)$:
  $J^P = \tfrac{1}{2}^{-}$ and $J^P = \tfrac{3}{2}^{-}$.
  The preferred interpretation within a contact-range effective field theory
  is that the $P_c(4440)$ is the $J^P = \tfrac{1}{2}^-$ molecule, while
  the $P_c(4457)$ is the $J^P = \tfrac{3}{2}^-$ one.
  Here we show that when the one pion exchange potential between
  the heavy-antimeson and heavy-baryon is taken into account,
  this conclusion changes, with the contrary identification
  being as likely as the original one.
  The identification is however cutoff dependent, which suggests
  that improvements of the present description (e.g. the inclusion of
  subleading order corrections, like two-pion exchanges) are necessary
  in order to disambiguate the spectroscopy of the molecular pentaquarks.
\end{abstract}

\pacs{13.60.Le, 12.39.Mk,13.25.Jx}

\maketitle
\section{Introduction}

The $P_c(4312)$, $P_c(4440)$ and $P_c(4457)$ are three hidden-charm
pentaquark-like states recently discovered
by the LHCb collaboration~\cite{Aaij:2019vzc}.
Owing to their closeness to the $\bar{D} \Sigma_c$ and $\bar{D}^* \Sigma_c$
thresholds, they have been theorized to be S-wave meson-baryon
bound states~\cite{Chen:2019bip,Chen:2019asm,Liu:2019tjn,Guo:2019fdo,Xiao:2019aya,Shimizu:2019ptd,Guo:2019kdc} (other explanations include
hadrocharmonium~\cite{Eides:2019tgv}
or a compact pentaquark~\cite{Wang:2019got,Cheng:2019obk}).
The most natural identification is that the $P_c(4312)$ is a
$\bar{D} \Sigma_c$ molecule and the $P_c(4440)$ and $P_c(4457)$ are
$\bar{D}^* \Sigma_c$ molecules.
This interpretation unambiguously predicts the quantum numbers of
the $P_c(4312)$ to be $J^P = \frac{1}{2}^-$.
In contrast there are two possibilities
for the quantum numbers of the $P_c(4440)$ and $P_c(4457)$:
$J^P = \frac{1}{2}^-$ and $J^P = \frac{3}{2}^-$.
That is, the identification is ambiguous.
Yet checking which quantum number corresponds to each one of
these two pentaquarks is important to clarify their nature,
in particular when confronted with future experimental
measurements of their properties.
From the recent theoretical models for the spectroscopy and decays of
these two molecules, the preferred identification so far
seems to be that the $P_c(4440)$ and $P_c(4457)$ are
the $J^P = \frac{1}{2}^-$ and $J^P = \frac{3}{2}^-$ $\bar{D}^* \Sigma_c$
molecules~\cite{Chen:2019asm,Liu:2019tjn,Xiao:2019aya},
respectively.
On the other hand, from the seminal predictions of molecular
hidden-charm pentaquarks we expect
the $J^P = \frac{1}{2}^-$ and $J^P = \frac{3}{2}^-$ $\bar{D}^* \Sigma_c$
molecules to be degenerate~\cite{Wu:2010jy,Wu:2010vk,Xiao:2013yca}
or for the $J^P = \frac{3}{2}^-$ $\bar{D}^* \Sigma_c$ state
to be the lighter than the $\frac{1}{2}^-$ one~\cite{Karliner:2015ina}.

The present manuscript considers this problem from the point of view of
spectroscopy within the effective field theory (EFT) framework.
Specifically we investigate the effect of including pion exchanges
in the masses of the $P_c(4312)$, $P_c(4440)$ and $P_c(4457)$ pentaquarks.
Previously Ref.~\cite{Liu:2018zzu} proposed a contact-range EFT to describe
the $\bar{D} \Sigma_c$ molecular states, which was used to predict
a $J^P = \tfrac{5}{2}^-$ $\bar{D}^* \Sigma_c^*$ molecular
pentaquark from the old $P_c(4450)$ peak~\cite{Aaij:2015tga}
(where we note that this state was first predicted
in Ref.~\cite{Xiao:2013yca}).
This EFT has been recently used in Ref.~\cite{Liu:2019tjn}
to analyze the LHCb pentaquark trio, where the following
two conclusions were reached:
(i) the molecular pentaquarks belong
to a multiplet with seven members (among which we count the aforementioned
$\tfrac{5}{2}^-$ state of Refs.~\cite{Xiao:2013yca,Liu:2018zzu}) and
(ii) the preferred quantum numbers for the $P_c(4440)$ and $P_c(4457)$ are
$J^P = \frac{1}{2}^-$ and $J^P = \frac{3}{2}^-$, respectively.
The first of these conclusions is relatively robust and has been independently
confirmed by other theoretical works~\cite{Xiao:2019aya,Sakai:2019qph,Yamaguchi:2019seo},
while the second is not so stringent,
as originally discussed in Ref.~\cite{Liu:2019tjn}.
Here we review these conclusions from the point of view of a pionful EFT,
i.e. a theory that besides contact-interactions also incorporates pions.
As we will see the inclusion of pions will be able to change the preferred
quantum number identification of the $P_c(4440)$ and $P_c(4457)$
pentaquarks (in agreement with the recent work of Ref.~\cite{Yamaguchi:2019seo}
which also considers the effects of pion exchanges).

The central idea of the present manuscript can be summarized as follows.
Heavy-quark spin symmetry (HQSS)~\cite{Isgur:1989vq,Isgur:1989ed}
when applied to hadronic molecules indicates that
the interaction among heavy hadrons is independent of
the spin of the heavy quarks within the aforementioned
heavy hadrons~\cite{Bondar:2011ev,Mehen:2011yh,Valderrama:2012jv,Nieves:2012tt,Lu:2017dvm}.
For the case of $\bar{D} \Sigma_c$ and $\bar{D}^* \Sigma_c$ molecules,
this symmetry implies that their S-wave potential takes
the form~\cite{Liu:2018zzu}
\begin{eqnarray}
  V(\bar{D}\Sigma_c, \tfrac{1}{2}) &=& V_a \, , \\
  V(\bar{D}^*\Sigma_c, \tfrac{1}{2}) &=& V_a - \frac{4}{3} V_b\, , \\
  V(\bar{D}^*\Sigma_c, \tfrac{3}{2}) &=& V_a + \frac{2}{3} V_b\, , 
\end{eqnarray}
with $V_a$ and $V_b$ a central and spin-spin contribution
that are in principle unknown.
If the particles are heavy enough we can assume that the binding energies
are proportional to the potential ($E \sim \langle V \rangle$),
in which case we find that the choice
\begin{eqnarray}
  E_a \sim -10 \,{\rm MeV} \enskip\, \mbox{and} \enskip\,
  \tfrac{2}{3} E_b \sim +5 \,{\rm MeV} \, , \label{eq:E-pionless}
\end{eqnarray}
indeed fits the spectrum of the pentaquark trio.
The inclusion of pion exchanges can potentially change this conclusion though.
One pion exchange (OPE) contains a spin-spin and a tensor piece: while the
spin-spin piece can be easily subsumed into the term $V_b$ of
the S-wave potential, the tensor piece will effectively generate
a central contribution to the $\bar{D}^* \Sigma_c$ molecules
that is not present in the $\bar{D} \Sigma_c$ system.
In practice we can modify the previous relations to
\begin{eqnarray}
  V(\bar{D}^*\Sigma_c, \tfrac{1}{2}) &=& V_a - \frac{4}{3} V_b + \delta V_a^{T}
  \, , \\
  V(\bar{D}^*\Sigma_c, \tfrac{3}{2}) &=& V_a + \frac{2}{3} V_b + \delta V_a^{T}
  \, , 
\end{eqnarray}
where $\delta V_a^{T}$ is the contribution to the tensor force~\footnote{This
  contribution is not necessarily the same in the spin-$\tfrac{1}{2}$ and
  -$\tfrac{3}{2}$ molecules (see Ref.~\cite{Yamaguchi:2019seo}),
  but the spin dependence can be reabsorbed in $V_b$
  leaving an effective tensor contribution
  which is spin independent.}.
If the effective contribution to the binding energy is
$\delta V_a^{T} \sim 5\,{\rm MeV}$, the preferred quantum numbers of the
pentaquark trio will change.
In fact the following identification
\begin{eqnarray}
  E_a \sim -10 \,{\rm MeV} \enskip\, ,  \enskip\,
  \tfrac{2}{3} E_b \sim -5 \,{\rm MeV}
  \enskip\, , \enskip\,
  \delta E^T_a \sim -5 \,{\rm MeV} , \nonumber \\
  \label{eq:E-pionful}
\end{eqnarray}
also fits the spectrum of the pentaquark trio.

However the previous is merely a heuristic argument which
has to be supported by concrete calculations.
HQSS for heavy hadron molecules does not directly apply to the binding energies,
but rather to the potential between heavy hadrons. 
As a consequence, HQSS will in general not translate into the type of
clean relations derived in the previous paragraph.
For instance, in analogy to the discussion around Eq.~(\ref{eq:E-pionless})
the predictions of pionless EFT prefers indeed the identification
of the $P_c(4440)$ with the $J^P = \tfrac{1}{2}^-$ $\bar{D}^* \Sigma_c$ molecule,
but there is room for the opposite identification
to be possible~\cite{Liu:2019tjn}.
In this manuscript we will investigate how the inclusion of pions modify
the previous conclusion.
In pionful EFT the opposite identification --- the $P_c(4440)$ is the
$J^P = \tfrac{3}{2}^-$ $\bar{D}^* \Sigma_c$ molecule --- is preferred,
yet the conclusion is not particularly strong at leading order.
Uncertainties both within pionless and pionful EFT make it not possible
to make a strong point based solely on spectroscopy.
Yet they suggest a preference.

The manuscript is organized as follows: in Sect.~\ref{sec:hqqs} we review
how HQSS applies to heavy baryon-meson molecules, in which we advocate
the use of a particular notation --- the light-quark
notation~\cite{Valderrama:2019sid} ---
for the description of the contact-range and the OPE potential within EFT.
In Sect.~\ref{sec:ope} we derive the one pion exchange potential
for the heavy antimeson-baryon system.
In Sect.~\ref{sec:Pc} we study the bound state spectrum for
the heavy antimeson-baryon system within the pionful EFT and
discuss their impact on the quantum numbers of
the known hidden-charm pentaquarks.
Finally, we present our conclusions in Sect.~\ref{sec:summary}.

\section{Heavy-Quark Spin Symmetry}
\label{sec:hqqs}

In this section we briefly explain how HQSS constrains
the interaction between a heavy meson and a heavy baryon.
For this, we will use two different notations.
The first is the standard heavy superfield notation,
in which we define a superfield that groups together
the heavy hadrons belonging to the same HQSS multiplet.
The second is the light-quark notation, which is based on the quark model
and in which we simply write down the light-quark subfield of
the heavy hadrons, see Ref.~\cite{Valderrama:2019sid} for a detailed
exposition and Refs.~\cite{Manohar:1992nd,Karliner:2015ina}
for previous examples of its use.

\subsection{Heavy Superfield Notation}

We begin by defining the superfields that are commonly used
for the description of heavy meson and heavy baryons.
The quark content of the S-wave heavy mesons is $Q \bar{q}$
with $Q$ and $q$ a heavy- and light-quark, respectively.
If the spin of the $Q \bar{q}$ pair couples to $S=0$ we have the ground
state heavy meson $P$ and if it couples to $S=1$ we have
the excited heavy meson $P^*$, where $P$ and $P^*$ are
degenerate in the limit in which the heavy-quark mass
goes to infinity.
For the $P$ and $P^*$ heavy mesons the non-relativistic superfield is
\begin{eqnarray}
  {H}_Q = \frac{1}{\sqrt{2}}\,
  \left[ P + \vec{P}^* \cdot \vec{\sigma} \right] \, ,
\end{eqnarray}
which is adapted from the relativistic
definition of Ref.~\cite{Falk:1992cx}.
$H_Q$ is a 2{$\times$}2 matrix and $\vec{\sigma}$ are the Pauli matrices.

For the S-wave heavy-baryons the quark content is $Q q q$.
If the light-quark pair is in the sextet configuration of
the SU(3)-flavor symmetry group (the case we will be considering here),
the spin of the light-quark pair couples to $S_L = 1$.
This implies that the total spin of the heavy-baryon is $S = \tfrac{1}{2}$
for the ground state $\Sigma_Q$ and $S = \tfrac{3}{2}$ for the excited
state $\Sigma_Q^*$, where $\Sigma_Q$ and $\Sigma_Q^*$ are degenerate
in the heavy-quark limit.
With this we define the non-relativistic superfield as
\begin{eqnarray}
 \vec{S}_Q = \frac{1}{\sqrt{3}}\,\vec{\sigma}\,\Sigma_Q + \vec{\Sigma}_Q^* \, ,
\end{eqnarray}
which again, corresponds to the non-relativistic limit of the superfield
originally defined in Ref.~\cite{Cho:1992cf}.

From the $H_{\bar Q}$ and $\vec{S}_Q$ superfields, the most general contact-range
Lagrangian with no derivatives we can construct is~\cite{Liu:2018zzu}
\begin{eqnarray}
  \mathcal{L} &=& 
  C_a \,
  \vec{S}_Q^{\dagger} \cdot \vec{S}_Q\,
      {\rm Tr}\left[ {\bar H}^{\dagger}_{\bar Q} {\bar H}_{\bar Q}\right]
  \nonumber \\
  &+& C_b\,\sum_{i = 1}^3\,
  \vec{S}_Q^{\dagger} \cdot (J_i \,\vec{S}_Q)\,
      {\rm Tr}\left[ {\bar H}^{\dagger}_{\bar Q} \sigma_i {\bar H}_{\bar Q}
        \right] \, ,
\end{eqnarray}
where $J_i$ with $i = 1,2,3$ refers to the spin-1 angular momentum matrices
and with $C_a$ and $C_b$ coupling constants.
Note that the $H_{\bar Q}$ superfield refers to the heavy-antimeson.
If we particularize for the $\bar{D} \Sigma_c$ family of molecules,
we obtain the contact-range potential of Table \ref{tab:VC-penta}.

\begin{table}[!ttt]
\begin{tabular}{|ccc|}
\hline \hline
  Molecule  & $J^{P}$ & $V_C$ \\
  \hline
  $\bar{D} \Sigma_c$ & $\frac{1}{2}^-$ & $C_a$ \\ \hline
  $\bar{D} \Sigma_c^*$ & $\frac{3}{2}^-$ & $C_a$ \\ \hline
  $\bar{D}^* \Sigma_c$ & $\frac{1}{2}^-$ & $C_a - \frac{4}{3}\,C_b$ \\
  $\bar{D}^* \Sigma_c$ & $\frac{3}{2}^-$ & $C_a + \frac{2}{3}\,C_b$ \\
  \hline
  $\bar{D}^* \Sigma_c^*$ & $\frac{1}{2}^-$ & $C_a - \frac{5}{3}\,C_b$ \\
  $\bar{D}^* \Sigma_c^*$ & $\frac{3}{2}^-$ & $C_a - \frac{2}{3}\,C_b$ \\
  $\bar{D}^* \Sigma_c^*$ & $\frac{5}{2}^-$ & $C_a + C_b$ \\
  \hline \hline 
\end{tabular}
\caption{The leading order contact-range potential
  for the charmed antimeson - charmed baryon system,
  i.e. the molecular hidden-charm pentaquarks.  
  We show the potential for each particle and spin channel
  (the ``Molecule'' and ``$J^P$'' columns),
  where the potential depends on two independent couplings $C_a$ and $C_b$.
  We do not explicitly show the isospin dependence of the couplings,
  but merely mention that the couplings in the $I=\tfrac{1}{2}$ and
  $\tfrac{3}{2}$ isospin configurations are different.
}
\label{tab:VC-penta}
\end{table}

\subsection{Light-Quark Notation}
\label{sec:light-quark}

Actually there is an easier and more direct method to write
the heavy-quark symmetric interactions, in which the idea
is to consider the heavy-quark as a spectator,
see Ref.~\cite{Valderrama:2019sid} for a detailed explanation.
Instead of writing superfields, we can write the interactions
in terms of the light-quark subfields.
For the $P$ and $P^*$ heavy mesons we consider the light-quark field
within the heavy mesons: $q_L$.
Equivalently, for the $\Sigma_Q$ and $\Sigma_Q^*$ heavy baryons
we use the light-diquark field within them: $d_L$.
With these $q_L$ and $d_L$ subfields,
the lowest order contact-range Lagrangian can be written as
\begin{eqnarray}
  \mathcal{L} &=& 
  C_a \, (q_L^{\dagger}\,q_L) \, (d_L^{\dagger}\,d_L)
  \nonumber \\
  &+& C_b\,(q_L^{\dagger}\,\vec{\sigma}_{L}\,q_L) \, \cdot \,
  (d_L^{\dagger}\,\vec{S}_{L}\,d_L) \, ,
\end{eqnarray}
where $\vec{\sigma}_L$ refers to the Pauli matrices as applied to
the $q_L$ field and $\vec{S}_L$ to the light-spin operators of
the $d_L$ field.
This Lagrangian leads to the contact-range potential
\begin{eqnarray}
  V(q_L\,d_L) = C_a + C_b\,\vec{\sigma}_{L1} \cdot \vec{S}_{L2} \, ,
  \label{eq:contact-light}
\end{eqnarray}
where the subscript $1$ and $2$ refer to the heavy meson and baryon,
respectively.
Now the contact-range potential is written in terms of the light-quark
subfields, i.e. in terms of the light-quark spin.
To rewrite the interactions in terms of the heavy hadron degrees of freedom
we apply a series of rules for translating the light-quark spin operators
into the heavy hadron spin operators.
For the heavy mesons the translation rules are
\begin{eqnarray}
  \langle P | \vec{\sigma}_L | P \rangle &=& 0 \, , \\
  \langle P^* | \vec{\sigma}_L | P^* \rangle &=& \vec{S}_1 \,  ,
\end{eqnarray}
where $\vec{S}_1$ refers to the spin-1 matrices as applied
to the heavy vector meson.
For the heavy baryons we have instead
\begin{eqnarray}
  \langle \Sigma_Q | \vec{J}_L | \Sigma_Q \rangle &=&
  \frac{2}{3}\,\vec{\sigma}_2 \, , \\
  \langle \Sigma_Q^* | \vec{J}_L | \Sigma_Q^* \rangle &=&
  \frac{2}{3}\,\vec{S}_2 \,  ,
\end{eqnarray}
where $\vec{\sigma}_2$ are the Pauli matrices
(applied to the spin-$\tfrac{1}{2}$ heavy baryon fields)
and $\vec{S}_2$ are the spin-$\tfrac{3}{2}$ angular momentum matrices
(applied to the spin-$\tfrac{3}{2}$ heavy baryon fields).
If we apply these substitution rules to the contact-range potential of
Eq.~(\ref{eq:contact-light}) for the light-quark subfields,
we arrive to the contact-range potential of Table \ref{tab:VC-penta}
written in the particle basis.
However the light-quark notation is much more compact and convenient,
as it reduces the seven possible heavy antimeson-baryon
potentials to a single formula.

\section{The One Pion Exchange Potential}
\label{sec:ope}

In this section we derive the OPE potential as applied
to the charmed antimeson and charmed baryon two-body system.
The derivation employs the light-quark notation presented
in Sect.~\ref{sec:light-quark}.
We discuss the coordinate and momentum space versions of the OPE potential
and its partial wave projection.

\subsection{Derivation of the Potential}

For the pion interactions, we begin by writing the following Lagrangians
written in terms of the superfields $H_Q$ and $\vec{S}_Q$:
\begin{eqnarray}
  \mathcal{L}_{HH \pi} &=&
  \frac{g_1}{\sqrt{2} f_{\pi}}\,
  {\rm Tr}[H^{\dagger}_{\bar Q} \, \tau_a \, \vec{\sigma}
    \cdot \vec{\nabla}\pi_a \, H_{\bar Q}] \, , \label{eq:L1} \\
  \mathcal{L}_{SS \pi} &=&
  \frac{i \, g_2}{\sqrt{2} f_{\pi}}\,\vec{S}^{\dagger}_Q \cdot
  ( T_a \, \vec{\nabla} \pi_a \times \vec{S}_Q) \, , \label{eq:L2}
\end{eqnarray}
with $g_1$, $g_2$ the axial couplings of the pion to the heavy meson
and heavy baryons, respectively, $f_{\pi} = 132\,{\rm MeV}$ the pion
decay constant, $\tau_a$ the Pauli matrices in isospin space,
$T_a$ the $I=1$ isospin matrices and where the latin index $a$
refers to the isospin.
For the axial couplings we choose
\begin{eqnarray}
  g_1 = 0.60 \quad \mbox{and} \quad g_2 = 0.84 \, ,
\end{eqnarray}
where $g_1$ is taken from the $D^* \to D \pi$ decays~\cite{Ahmed:2001xc,Anastassov:2001cw}
($g_1 = 0.59 \pm 0.01 \pm 0.07$)
and $g_2$ from the lattice QCD calculation of Ref.~\cite{Detmold:2012ge}.
We notice that there are several conventions for $g_2$, which are
discussed in Ref.~\cite{Detmold:2012ge} (from which one can also
find the relations among them).
The convention we use here differs by a sign of the one
by Cho~\cite{Cho:1992cf}, i.e. $g_2 = - g_{2,\rm Cho}$.
From this Lagrangian we can write the OPE potential as
\begin{eqnarray}
  V_{\rm OPE} =
  \frac{\mathcal{A}_1(\vec{q}) \mathcal{A}_2(-\vec{q})}{q^2 + m_{\pi}^2} \, ,
\end{eqnarray}
where $\mathcal{A}_1$ and $\mathcal{A}_2$ refer to the non-relativistic
amplitudes
\begin{eqnarray}
  \mathcal{A}_1 &=& \mathcal{A}(H_Q \to H_Q' \pi) \, , \\
  \mathcal{A}_2 &=& \mathcal{A}(S_Q \to S_Q' \pi) \, ,
\end{eqnarray}
in the non-relativistic normalization of the amplitudes
used in Refs.~\cite{Valderrama:2012jv,Lu:2017dvm}
(but notice that Ref.~\cite{Lu:2017dvm} uses the normalization of
Cho~\cite{Cho:1992cf} for the axial coupling of the heavy baryon).
By specifying $\mathcal{A}_1$ and $\mathcal{A}_2$ for the particular
heavy meson and heavy baryon of interest,
we can obtain the potential
for any of the cases.
The procedure ends in seven possible potentials,
one for each of the possible S-wave molecules,
which we will not write here in detail.

Alternatively, we can write the Lagrangians of Eqs.~(\ref{eq:L1})
and (\ref{eq:L2}) in terms of the light-quark fields
within the heavy hadrons:
\begin{eqnarray}
  \mathcal{L}_{q_L q_L \pi} &=& \frac{g_1}{\sqrt{2} f_{\pi}}\,
        {q}_{L}^{\dagger}
        \vec{\sigma}_{L} \cdot \vec{\nabla} ( {\tau}_a \, {\pi}_a)
        q_{L} \, , \\
  \mathcal{L}_{d_L d_L \, \pi} &=& \frac{g_2}{\sqrt{2} f_{\pi}}\,
        {d}_{L}^{\dagger}
        \vec{S}_{L} \cdot \vec{\nabla} ( {T}_a \, {\pi}_a)
        d_{L} \, . 
\end{eqnarray}
From this, the OPE potential can be written in momentum space as
\begin{eqnarray}
  V_{\rm OPE}(\vec{q}) = -\frac{g_1 g_2}{2 f_{\pi}^2}\,
  \vec{\tau}_1 \cdot \vec{T}_2\,
  \frac{\vec{\sigma}_{L1} \cdot \vec{q} \,\vec{S}_{L2} \cdot \vec{q}}
       {\vec{q}^2 + m_{\pi}^2} \, .
\end{eqnarray}
We can Fourier-transform the OPE potential into coordinate space
\begin{eqnarray}
  V_{\rm OPE}(\vec{r}) &=& -\frac{g_1 g_2}{6 f_{\pi}^2}\,
  \vec{\tau}_1 \cdot \vec{T}_2\, \vec{\sigma}_{L1} \cdot \vec{S}_{L2}\,
  \delta^{(3)}(\vec{r}) \\
  &+& \vec{\tau}_1 \cdot \vec{T}_2\,\left[
    \vec{\sigma}_{L1} \cdot \vec{S}_{L2}\,W_C(r) +
    S_{L12}(\hat{r)}\,W_C(r) \right] \, , \nonumber \\
\end{eqnarray}
where $W_C$ and $W_T$ are defined as
\begin{eqnarray}
  W_C(r) &=& \frac{g_1 g_2 m_{\pi}^3}{24 \pi f_{\pi}^2}\,
  \frac{e^{-m_{\pi} r}}{m_{\pi} r} \, , \\
  W_T(r) &=& \frac{g_1 g_2 m_{\pi}^3}{24 \pi f_{\pi}^2}\,
  \frac{e^{-m_{\pi} r}}{m_{\pi} r}\,\left(1 + \frac{3}{m_{\pi} r} +
  \frac{3}{(m_{\pi} r)^2}  \right) \, .
\end{eqnarray}

\subsection{Partial Wave Projection}

Strong interactions preserve the total angular momentum
$\vec{J}= \vec{L} + \vec{S}$, but not the orbital
angular momentum or spin separately.
As a consequence the OPE potential will mix partial waves with the same quantum
number $J$, but different quantum numbers $L$ and $S$.
If we use the spectroscopic notation $^{2S+1}L_J$, the partial waves
comprising the three pentaquark-like $\bar{D} \Sigma_c$ and
$\bar{D}^* \Sigma_c$ molecular candidates are
\begin{eqnarray}
  | \bar{D} \Sigma_c(\tfrac{1}{2}^-) \rangle &=& \{ ^2S_{\tfrac{1}{2}} \} \, , \\
  \nonumber \\
  | \bar{D}^* \Sigma_c(\tfrac{1}{2}^-) \rangle &=&
  \{ ^2S_{\tfrac{1}{2}}, ^4D_{\tfrac{1}{2}} \} \, , \\
  | \bar{D}^* \Sigma_c(\tfrac{3}{2}^-) \rangle &=&
  \{ ^2D_{\tfrac{3}{2}}, ^4S_{\tfrac{3}{2}}, ^4D_{\tfrac{3}{2}} \} \, ,
\end{eqnarray}
plus the corresponding decomposition for the other four $\bar{D} \Sigma_c^*$
and $\bar{D}^* \Sigma_c^*$ molecular configurations containing S-waves.

The partial wave projection is done by defining
a generalized spherical harmonic for the $^{2S+1}L_J$ wave
\begin{eqnarray}
  \mathcal{Y}^{LS}_{JM}(\Omega) = \sum_{M_L M_S} Y_{L M_L}(\Omega) | S M_S \rangle
  \langle L M_L S M_S | J M \rangle \, ,
\end{eqnarray}
where $\Omega$ is the solid angle and which can be used to project the
potential into the partial wave basis.
For the momentum space potential this is done as follows
\begin{eqnarray}
  && \langle k, J LS | V | k', J L'S' \rangle = \nonumber
  \\ && \qquad \frac{1}{4\pi}\,
  \int d\hat{k} d\hat{k}' \, \mathcal{Y}^{LS*}_{JM}(\hat{k})
  \,V(\vec{k}' - \vec{k})\,
  \mathcal{Y}^{L'S'}_{JM}(\hat{k}') \, ,
\end{eqnarray}
while for the coordinate space potential we have
\begin{eqnarray}
  && \langle J LS | V(r) | J L'S' \rangle = \nonumber
  \\ && \qquad 
  \int d\hat{r} \, \mathcal{Y}^{LS*}_{JM}(\hat{r})\, V(\vec{r}) \,
  \mathcal{Y}^{L'S'}_{JM}(\hat{r}) \, , \label{eq:pw-r}
\end{eqnarray}
where the projection is independent of the third component of
the total angular momentum $M$.
In coordinate space a further simplification is possible by noticing that
the partial wave projection only involves writing the spin-spin and
tensor operators as matrices in the space of the partial waves
comprising a particular state:
\begin{eqnarray}
  \vec{\sigma}_{L1} \cdot \vec{S}_{L2} \to {f_{12}}\,{\bf C}_{12}
  \quad \mbox{and} \quad
  S_{L12} \to {f_{12}}\,{\bf S}_{12} \, , 
\end{eqnarray}
where $f_{12}$ is a conversion factor ($f_{12} = \frac{2}{3}$ in all cases)
and the matrices ${\bf C}_{12}$ and ${\bf S}_{12}$
can be consulted in Table \ref{tab:operators}.

\section{The Molecular Pentaquark Spectrum}
\label{sec:Pc}

In this section we discuss the description of the LHCb pentaquark trio
--- the $P_c(4312)$, $P_c(4440)$ and $P_c(4457)$ --- within
the molecular picture in a pionful EFT.
We will consider the $P_c(4312)$ as a $\bar{D} \Sigma_c$ bound state and
the $P_c(4440)$ and $P_c(4457)$ as $\bar{D}^* \Sigma_c$ ones.
The consistent description of the pentaquark trio suggest a slight
preference for the quantum numbers $J^P = \tfrac{3}{2}^-$
and $\tfrac{1}{2}^-$ for the $P_c(4440)$ and $P_c(4457)$,
respectively.
The pionful EFT will also lead to the prediction of
other four molecular pentaquarks.

\begin{table}[!ttt]
\begin{tabular}{|cccc|}
\hline \hline
  Scenario & $\Lambda$ $({\rm MeV})$ & $C_a$ $({\rm fm}^2)$ & $C_b$ $({\rm fm}^2)$   \\
  \hline
  $A$ & $0.75$ & $-1.1199$ & $-0.1183$  \\ 
  $A$ & $1.50$ & $-0.3466$ & $-0.1669$  \\ 
  \hline
  $B$ & $0.75$ & $-1.2755$ & $-0.5494$  \\ 
  $B$ & $1.50$ & $-0.4001$ & $-0.3760$  \\ 
  \hline \hline
  Scenario & $R_c$ $({\rm fm})$  & $C_a$ $({\rm fm}^2)$ & $C_b$ $({\rm fm}^2)$  \\
  \hline
  $A$ & $0.5$ & $-0.5741$ & $+0.1345$  \\ 
  $A$ & $1.0$ & $-1.7447$ & $+0.4074$  \\ 
  \hline
  $B$ & $0.5$ & $-0.6494$ & $-0.0400$  \\ 
  $B$ & $1.0$ & $-2.0142$ & $-0.3503$  \\ 
  \hline \hline
\end{tabular}
\caption{
  The contact-range couplings $C_a$ and $C_b$ from the condition of reproducing
  the mass of the $P_c(4440)$ and $P_c(4457)$ as molecular pentaquarks
  in p- and r-space (as indicated by type of cutoff: $\Lambda$ and $R_c$).
  Scenario A corresponds to considering that the spin-parities of
  the $P_c(4440)$ and $P_c(4457)$ are $J^P = \tfrac{1}{2}^-$ and
  $\tfrac{3}{2}^-$, respectively, while scenario B corresponds to the opposite
  identification.
}
\label{tab:couplings}
\end{table}

\begin{table}[!ttt]
\begin{tabular}{|ccccc|}
\hline \hline
  Scenario & Molecule  & $J^{P}$ & B (MeV) & M (MeV) \\
  \hline
  $A$ & $\bar{D} \Sigma_c$ & $\frac{1}{2}^-$ &
  $(2)^V-7$ & $4314-(4319)^V$ \\ \hline
  $A$ & $\bar{D} \Sigma_c^*$ & $\frac{3}{2}^-$ &
  $(1)^V-7$ & $4378-(4384)^V$  \\ \hline
  $A$ & $\bar{D}^* \Sigma_c$ & $\frac{1}{2}^-$ & Input &$4440.3$ \\
  $A$ & $\bar{D}^* \Sigma_c$ & $\frac{3}{2}^-$ 
  & Input & $4457.3$ \\
  \hline
  $A$ & $\bar{D}^* \Sigma_c^*$ & $\frac{1}{2}^-$ 
  & $27-44$ & $4483-4500$
  \\
  $A$ & $\bar{D}^* \Sigma_c^*$ & $\frac{3}{2}^-$ & 
  $16-20$ & $4507-4512$
  \\
  $A$ & $\bar{D}^* \Sigma_c^*$ & $\frac{5}{2}^-$ &
  $4-6$ & $4520-4523$ \\
  \hline \hline 
  $B$ & $\bar{D} \Sigma_c$ & $\frac{1}{2}^-$ & $0-12$ &
  $4308-4321$\\ \hline
  $B$ & $\bar{D} \Sigma_c^*$ & $\frac{3}{2}^-$ &
  $0-13$ & $4372-4385$ 
   \\ \hline
  $B$ & $\bar{D}^* \Sigma_c$ & $\frac{1}{2}^-$  & Input & $4457.3$ \\
  $B$ & $\bar{D}^* \Sigma_c$ & $\frac{3}{2}^-$  & Input &$4440.3$ \\
  \hline
  $B$ & $\bar{D}^* \Sigma_c^*$ & $\frac{1}{2}^-$ &
  $4-14$ & $4513-4523$
  \\
  $B$ & $\bar{D}^* \Sigma_c^*$ & $\frac{3}{2}^-$ &
  $11-16$ & $4511-4516$
  \\
  $B$ & $\bar{D}^* \Sigma_c^*$ & $\frac{5}{2}^-$ &
  $26-29$ & $4497-4501$ \\
  \hline \hline
\end{tabular}
\caption{
  Predictions for the S-wave HQSS molecular multiplet of
  heavy antimeson-baryon molecules, as derived from the 
  lowest{-}order potential in pionful EFT (p-space).
  This potential contains a contact-range piece with two unknown
  couplings $C_a$ and $C_b$ and a finite-range piece, given by OPE.
  In all cases we assume that the isospin of the listed
  molecules is $I=\tfrac{1}{2}$.
  We determine the value of the $C_a$ and $C_b$ couplings from the condition
  of reproducing the location of the $P_c(4440)$ and $P_c(4457)$ resonances,
  which are known to be close to the $\bar{D}^* \Sigma_c$ threshold.
  We do not know however the quantum numbers of the $P_c(4440)$ and $P_c(4457)$,
  but consider two possibilities instead, scenario $A$ and $B$,
  where in the first the $P_c(4440)$ is the $\frac{1}{2}^-$ molecule
  and in the second the $P_c(4457)$ is the $\frac{1}{2}^-$ molecule.
  If a molecular pentaquark becomes unbound but survives as a virtual state
  (a situation that happens for the $\bar{D} \Sigma_c$ and
  $\bar{D} \Sigma_c^*$ systems),
  we indicate this situation with the superscript $V$.
  Calculations are done in momentum space with the regularization described
  in Eq.~(\ref{eq:V-reg}) and a cutoff $\Lambda = 0.75-1.5\,{\rm GeV}$.
}
\label{tab:penta-p}
\end{table}

\begin{table}[!ttt]
\begin{tabular}{|ccccc|}
\hline \hline
  Scenario & Molecule  & $J^{P}$ & B (MeV) & M (MeV) \\
  \hline
  $A$ & $\bar{D} \Sigma_c$ & $\frac{1}{2}^-$ & $1-8$ & $4313-4320$ \\ \hline
  $A$ & $\bar{D} \Sigma_c^*$ & $\frac{3}{2}^-$ & $1-8$ & $4377-4384$ \\ \hline
  $A$ & $\bar{D}^* \Sigma_c$ & $\frac{1}{2}^-$ & Input &$4440.3$ \\
  $A$ & $\bar{D}^* \Sigma_c$ & $\frac{3}{2}^-$ 
  & Input & $4457.3$ \\
  \hline
  $A$ & $\bar{D}^* \Sigma_c^*$ & $\frac{1}{2}^-$ 
  & $28-36$ & $4490-4499$
  \\
  $A$ & $\bar{D}^* \Sigma_c^*$ & $\frac{3}{2}^-$ & 
  $17-20$ & $4507-4510$
  \\
  $A$ & $\bar{D}^* \Sigma_c^*$ & $\frac{5}{2}^-$ &
  $4-7$ & $4520-4523$ \\
  \hline \hline 
  $B$ & $\bar{D} \Sigma_c$ & $\frac{1}{2}^-$ & $5-14$ &
  $4307-4315$\\ \hline
  $B$ & $\bar{D} \Sigma_c^*$ & $\frac{3}{2}^-$ & $6-14$ &
  $4371-4379$ \\ \hline
  $B$ & $\bar{D}^* \Sigma_c$ & $\frac{1}{2}^-$  & Input & $4457.3$ \\
  $B$ & $\bar{D}^* \Sigma_c$ & $\frac{3}{2}^-$  & Input &$4440.3$ \\
  \hline
  $B$ & $\bar{D}^* \Sigma_c^*$ & $\frac{1}{2}^-$ &
  $3-8$ & $4518-4523$
  \\
  $B$ & $\bar{D}^* \Sigma_c^*$ & $\frac{3}{2}^-$ &
  $11-15$ & $4512-4516$
  \\
  $B$ & $\bar{D}^* \Sigma_c^*$ & $\frac{5}{2}^-$ &
  $28-33$ & $4494-4499$ \\
  \hline \hline
\end{tabular}
\caption{
  Predictions for the S-wave HQSS molecular multiplet of
  heavy antimeson-baryon molecules, as derived from the 
  lowest{-}order potential in pionful EFT (r-space).
  We refer to Table \ref{tab:penta-p} for details.
  Calculations are done in coordinate space with the regularization described
  in Eqs.~(\ref{eq:VC-reg}) and (\ref{eq:VF-reg}) and a cutoff
  $R_c = 0.5-1.0\,{\rm fm}$.
  The G-wave components are ignored for the $J^P = \tfrac{3}{2}^-$
  and $\tfrac{5}{2}^-$ $\bar{D}^* \Sigma_c^*$ molecules,
  as their contribution to the binding energy
  is negligible.
}
\label{tab:penta-r}
\end{table}

\subsection{Bound state equations}

We calculate the binding energies of a heavy baryon-antibaryon bound state
by plugging the EFT potential into the Lippmann-Schwinger or Schr\"odinger
equation, depending on whether the EFT potential has been written
in momentum or coordinate space.
For momentum space, the bound state equation takes the form
\begin{eqnarray}
  \phi_{LS}^J(p) &=& \sum_{L'S'} \int \frac{d^3 q}{(2\pi)^3}
  \frac{\langle p, J LS | V | q, J L'S'\rangle}{E - \frac{q^2}{2 \mu}}\,
  \phi_{L'S'}^J(q)
  \, ,
\end{eqnarray}
where $L$, $S$ and $J$ are the orbital, intrinsic and total angular momentum,
with $\phi_{LS}^J$ the vertex function.
%
This bound state equation can be solved by discretizing this integral
equation and finding the eigenvalues of the ensuing linear equations.
For coordinate space, we use the reduced Schr\"odinger equation
\begin{eqnarray}
  -{u_{LS}^J}'' + 2 \mu \sum_{L' S'} \, V_{LS, L'S'}^J(r)\,u_{L'S'}^J(r)
  && \nonumber \\
  +
  \frac{L(L+1)}{r^2} \, u_{L S}^J(r) &=&
  -\gamma^2\,u_{L S}^J(r) \, , \nonumber \\
\end{eqnarray}
which is a system of coupled ordinary differential equations
that can be solved by standard means.
%

\subsection{Regularization and renormalization}

The EFT potential is not well-behaved at distances below the pion Compton
wavelength, a problem that is taken care of by means of a
regularization and renormalization procedure.
The regularization part is as follows: for the momentum space version of
the potential, we use a separable regulator of the type
\begin{eqnarray}
  \langle p' | V_{\Lambda} | p \rangle  = \langle p'  | V | p \rangle\,
  f(\frac{p'}{\Lambda})\,f(\frac{p}{\Lambda}) \, ,
  \label{eq:V-reg}
\end{eqnarray}
where $f(x) = e^{-x^2}$, i.e. a Gaussian regulator.
For the coordinate space potential we use a local regulator,
which is different depending on whether it is a applied for the contact-
or finite-range piece of the EFT potential.
For the regularization of the contact-range potential,
we use a Gaussian regulator of the type
\begin{eqnarray}
  \delta^{(3)}(\vec{r}) \to
  \frac{e^{-(r/R_c)^4}}{\frac{4}{3} \pi \Gamma(\frac{7}{4}) R_c^3} \, ,
  \label{eq:VC-reg}
\end{eqnarray}
while for the OPE potential we use
\begin{eqnarray}
  V_{\rm OPE}(r) \to V_{\rm OPE}(r) \, \left[ 1 - e^{-(r/R_c)^4} \right] \, .
  \label{eq:VF-reg}
\end{eqnarray}
This type of local r-space regulators have been recently put in use
in pionful EFT as applied to nuclear physics~\cite{Gezerlis:2014zia}.
We choose the Gaussian exponent to be $n=4$ as this is enough
to suppress the divergence of the tensor force
at short distances.

For the renormalization part, the idea is that the contact-range couplings,
$C_a$ and $C_b$ in this case, will be able to absorb the cutoff dependence.
Thus the predictions derived within the EFT framework are
expected to be cutoff independent.
For checking the cutoff independence hypothesis, we choose the following
cutoff window in momentum space
\begin{eqnarray}
  \Lambda = (0.75-1.5)\,{\rm GeV} \, ,
\end{eqnarray}
which roughly corresponds to $\{ m_{\rho}, 2 m_{\rho} \}$.
This window is harder than the one we previously used in the contact-range
EFT of Ref.~\cite{Liu:2019tjn}, i.e. $\Lambda = 0.5-1.0\,{\rm GeV}$.
The choice of a harder cutoff is driven by the experience
from pionful EFT as applied to heavy meson-antimeson
molecules~\cite{Baru:2016iwj,Baru:2017gwo},
in which larger cutoffs than in a purely contact theory
seemed to make a difference.
For the coordinate space calculation we choose
\begin{eqnarray}
  R_c = 0.5-1.0\,{\rm fm} \, ,
\end{eqnarray}
which comes from rounding up the $\{ \pi / 2 m_{\rho} , \pi / m_{\rho} \}$
cutoff window. This is approximately equivalent to the momentum space
window if we consider the relation $R_c = \pi / \Lambda$
for the r- and p-space cutoffs.
Unfortunately cutoff independence is not achieved at the accuracy level
we will require to unambiguously distinguish the quantum numbers
of the $P_c$ pentaquarks.

\subsection{The quantum numbers of the pentaquark trio}

The couplings $C_a$ and $C_b$ are actually determined from observable
quantities, for which we will use the binding energies of
the $P_c(4440)$ and $P_c(4457)$ pentaquarks.
The natural expectation in the molecular picture is that the $P_c(4440)$
and $P_c(4457)$ are $\bar{D}^* \Sigma_c$ bound states with isospin
$I=\tfrac{1}{2}$, for which two possibilities exist
for the total angular momentum: $J=\tfrac{1}{2}$ and $J=\tfrac{3}{2}$.
We do not know which is the total angular momentum of each of the molecular
pentaquark candidates, which means that we will consider two scenarios:
\begin{itemize}
\item[(i)] scenario A: the $P_c(4440)$ is the $J = \tfrac{1}{2}$ molecule \\
  (while the $P_c(4457)$ is the $J = \tfrac{3}{2}$ molecule),
\item[(ii)] scenario B: the $P_c(4440)$ is the $J = \tfrac{3}{2}$ molecule\\
  (while the $P_c(4457)$ is the $J = \tfrac{1}{2}$ molecule),
\end{itemize}
which are the same two scenarios considered in Ref.~\cite{Liu:2019tjn}.
The values of the couplings $C_a$ and $C_b$ that are obtained
in each scenario can be consulted in Table \ref{tab:couplings}.
Each of the scenarios predicts a different mass for the $P_c(4312)$ pentaquark.
In momentum space, scenario $A$ predicts
\begin{eqnarray}
  M_1^A = 4314-(4319)^V\,{\rm MeV} \, ,
\end{eqnarray}
where the only uncertainty we have taken into account is the cutoff variation,
with the $V$ superscript standing for the fact that the bound state
disappears and becomes a virtual state instead
for $\Lambda = 1.5\,{\rm GeV}$. 
On the other hand scenario $B$ predicts
\begin{eqnarray}
  M_1^B = 4308-4321\,{\rm MeV} \, .
\end{eqnarray}
This preliminary comparison indicates that scenario $B$ is slightly favored
over scenario $A$, but the conclusion is merely tentative at best.

The residual cutoff variation alone already indicates that the error of
the pionful EFT at leading order is probably too large to distinguish
between the two scenarios.
Besides the cutoff uncertainty, there are two other error sources
that we have not explicitly considered:
the uncertainty (i) in HQSS and (ii) in the $g_2$ axial coupling
constant of the pion with the sextet heavy baryons.
Regarding (i), HQSS, the location of the $P_c(4312)$ is determined from the
contact-range coupling $C_a$, but in doing so we are assuming
that HQSS is exact for the hidden-charm molecular pentaquarks.
This is not the case, with HQSS violations expected to have a size of
$\Lambda_{\rm QCD} / m_c$, with $\Lambda_{\rm QCD} \sim 200-300\,{\rm MeV}$
and $m_c$ the charm quark mass, yielding a $15-20\%$ variation
for the coupling $C_a$ around the determination we have done.
Regarding (ii), the $g_2$ axial coupling, the uncertainty
in the lattice QCD calculation is sizable: $g_2 = 0.84 \pm 0.20$.
Besides, this lattice QCD calculation applies to the heavy-quark limit
($m_Q \to \infty$, with $m_Q$ the mass of the heavy quark).
The $g_2$ axial coupling can be derived from the axial coupling involved
in the sextet to antitriplet heavy baryon transitions, $g_3$, and
a quark model relation (see Ref.~\cite{Cheng:2015naa}
for a comprehensive review, which uses the normalization
of Yan~\cite{Yan:1992gz} for the axial couplings).
In turn the $g_3$ axial coupling can be determined
from the $\Sigma_c \to \Lambda_c \pi$ decay.
This procedure yields $g_2 \sim 1.4$~\cite{Cheng:2015naa},
a value considerably larger than the one we have chosen
(and which indeed makes a difference).
If this were not enough, the location of the $P_c(4312)$ is not known
with the required accuracy either.
A recent theoretical exploration has proposed that the $P_c(4312)$ is
a virtual state instead of a bound state~\cite{Fernandez-Ramirez:2019koa}:
if this is the case, scenario $A$ should be the preferred one.

We recognize the following three factors influencing
the preference over scenarios $A$ and $B$:
\begin{itemize}
\item[(i)] softer cutoffs ($\Lambda \sim 0.5\,{\rm GeV}$) 
  favor scenario $A$, while harder ones ($\Lambda \gtrsim 1\,{\rm GeV}$)
  favor scenario $B$,
\item[(ii)] larger axial couplings ($g_2 \sim 1.4$) favor scenario $B$,
\item[(iii)] a less bound (or virtual) $P_c(4312)$ favors scenario $A$.
\end{itemize}
The first of these factors refers to the inner workings of the EFT and probably
can be only dealt with by improving the current EFT description,
e.g. calculating the subleading order corrections~\footnote{
  We notice in passing that the subleading EFT potential has been calculated
  in Ref.~\cite{Meng:2019ilv}, though with the aim of deducing the existence
  of the pentaquark trio from the two-nucleon system (by extrapolating
  the contact-range couplings from the two-nucleon system to
  the heavy antimeson-baryon system). That is, the use of pionful EFT
  in Ref.~\cite{Meng:2019ilv} is very different from the one
  in the present manuscript. Nonetheless we point out
  that it might be possible to combine the subleading potential
  of Ref.~\cite{Meng:2019ilv} with the ideas of Ref.~\cite{Baru:2017gwo}
  (properly adapted from the heavy meson-antimeson to
  the heavy baryon-antimeson case) to better pinpoint
  the quantum numbers of the pentaquark trio.},
which will require new data as the next-to-leading order
contact-range potential will involve new couplings.
The second of these factors is difficult to settle experimentally --- the $g_2$
axial coupling does not directly appear in decays or other quantities that
are directly observable~\cite{Cheng:2015naa} --- but can probably
be determined by lattice QCD calculations that take into account
the finite charm quark mass.
The third factor can eventually be determined in future experiments
with smaller uncertainties.

At this point it is important to comment about cutoff independence.
In principle we expect cutoff independence to be achieved by means of
the renormalization process, where the contact-range couplings
--- $C_a$ and $C_b$ in this case --- are expected to absorb
the divergences associated with the short-range
quirks of the EFT potential.
However this is not the case for the calculations presented here:
the effects of the tensor force have not been completely
reabsorbed in the couplings $C_a$ and $C_b$.
The manifestation of this problem is the binding energy prediction of the 
$P_c(4312)$ pentaquark.
If we assume it to be a $\bar{D} \Sigma_c$ molecule this system cannot
exchange pions.
As the cutoff $\Lambda$ grows, the effect of the tensor force
will be increasingly attractive, forcing the $C_a$ coupling to
be less and less attractive.
Eventually, for $\Lambda$ hard enough, the $P_c(4312)$ will cease to be bound
and will become a virtual state instead.
In momentum space this indeed happens for scenario $A$ and a cutoff
of the order of $1.5\,{\rm GeV}$. It also happens for scenario $B$,
though in this case a harder cutoff is required
(around $2.0\,{\rm GeV}$, give or take).

This is bad news because it partially invalidates one of the expected
advantages of the EFT framework over phenomenological models:
systematic error estimations.
In a properly renormalized EFT, where calculations do not strongly depend
in the cutoff, the cutoff variation might be used as a proxy of
the EFT uncertainty.
However it is impossible to describe the LHCb pentaquark trio in a cutoff
independent way: large cutoffs invariably lead to the disappearance of
the $P_c(4312)$ member of the trio.
Of course this happens for relatively hard cutoffs
in the $1.5-2.0\,{\rm GeV}$ range, which means that
this disappearance is not physically relevant
but rather an artifact.
Yet, despite being an artifact,
it prevents the systematic estimation of the theoretical uncertainty.
Basically, even if the experimental error in the determination of
the $P_c(4312)$ mass was negligible, there will be no completely
model independent way to distinguish both scenarios
in the pionful EFT proposed here.
Despite this drawback, pionful calculations are still useful even if
they begin to show a sizable cutoff dependence
at $\Lambda > 1.5\,{\rm GeV}$.
It is interesting to notice that a similar cutoff dependency
has been discussed for EFTs involving heavy flavor
symmetry~\cite{Baru:2018qkb},
which is a different manifestation of heavy quark symmetry.
Be it as it may, the degree of model dependence is probably
smaller than for phenomenological models.

The conclusion is that there is a preference for scenario B.
The fact that this preference is not particularly strong is in line with the
early speculations about the existence of molecular pentaquarks,
in which predictions showed a clear degeneracy
in spin~\cite{Wu:2010jy,Wu:2010vk,Xiao:2013yca}.
The inclusion of pions simply points towards the hypothesis of Karliner
and Rosner~\cite{Karliner:2015ina}, where the $J^P = \tfrac{3}{2}^-$
molecular pentaquark is expected to be more bound
than its $J^P = \tfrac{1}{2}^-$ partner.
In contrast in the traditional one boson exchange model
this pattern is apparently inverted~\cite{Chen:2019bip,Cheng:2019obk},
with the lower spin molecules being more bound
than the higher spin ones.
However a recent work~\cite{Liu:2019-recent},
which has revisited the application of the one boson exchange model
to heavy antimeson-baryon molecules, suggests that
this is not necessary the case and that
scenario B might be the most probable.

\subsection{The pentaquark HQSS septuplet}

The consistent description of the $P_c(4312)$, $P_c(4440)$ and $P_c(4457)$
pentaquark trio in the molecular picture fully determines
the LO potential in pionful EFT.
As a consequence we can compute the binding energies of all the S-wave
molecular configurations.
The results are summarized in Tables \ref{tab:penta-p} and \ref{tab:penta-r}
for the momentum and coordinate space versions~\footnote{
  The momentum space calculation contains all the partial waves,
  including the G-waves in the $J^P = \tfrac{3}{2}^-$ and
  $\tfrac{5}{2}^-$ $\bar{D}^* \Sigma_c^*$ molecules
  (see Table \ref{tab:operators}) ,
  the contribution of which can be checked
  to be negligible (less than $0.1\,{\rm MeV}$).
  In view of this result, the coordinate space
  calculation ignores the G-waves,
  which greatly simplifies the required computations.
} of the LO potential.
As happened in the pionless EFT at LO~\cite{Liu:2019tjn},
we predict the seven possible HQSS partners of the pentaquark trio,
independently of whether we use scenario A or B
for the $P_c(4440)$ and $P_c(4457)$ quantum numbers.
The most important difference with the contact-range theory is
that the predictions for the $\bar{D} \Sigma_c$ and
$\bar{D} \Sigma_c^*$ molecules are less bound,
leading to a marginal preference of
scenario B over A.
In every other respect, Tables \ref{tab:penta-p} and \ref{tab:penta-r}
only confirm the patterns already discovered in Ref.~\cite{Liu:2019tjn}:
scenario A (B) leads to the higher spin states
being more (less) massive.
If this were not enough, further confirmation can be found in the recent
pionless EFT calculation of Ref.~\cite{Sakai:2019qph}, which also considers
transitions among the $\bar{D} \Sigma_c$ , $\bar{D} \Sigma_c^*$,
$\bar{D}^* \Sigma_c$ and $\bar{D}^* \Sigma_c^*$ channels.
In this regard, the eventual discovery of a
$\tfrac{5}{2}^-$ $\bar{D}^* \Sigma_c^*$ molecule
will probably settle the question about the quantum numbers of the 
$P_c(4440)$ and $P_c(4457)$: the prediction of the location of
this $\bar{D}^* \Sigma_c^*$ molecules varies by about $20-25\,{\rm MeV}$
depending on the scenario.
However, owing to its angular momentum $J = \tfrac{5}{2}$, the experimental
detection of a $\bar{D}^* \Sigma_c^*$ pentaquark state is not probable
in the $J/\Psi p$ channel where the other pentaquarks
have been discovered.
The $J = \tfrac{5}{2}$ state might indeed be difficult to observe
from its decays to a charmonium: all possible charmonium decays
for this state are p- or d-wave, which indicates
that they might be relatively suppressed.

Notice that other works lead to different predictions of the septuplet.
In Ref.~\cite{Xiao:2019aya} the binding energy of the molecular
pentaquarks is almost independent of the spin and
the identification between scenarios A and B is done
on the basis of the predicted decay widths.
This approximate degeneracy of the binding energy is however
a consequence of explicitly ignoring the coupling $C_b$:
Ref.~\cite{Xiao:2019aya} determines the couplings from resonance
saturation in the hidden gauge model, with $C_b$ receiving its main
contribution from OPE, which is assumed to be weak.
Ref.~\cite{Shimizu:2019ptd} also predicts a multiplet structure
for the hidden charm pentaquarks, which relies on HQSS and OPE.
But the multiplet structure of Ref.~\cite{Shimizu:2019ptd}
is merely a subset of the septuplet of
Refs.~\cite{Liu:2019tjn,Xiao:2019aya}.
The reason for the difference is that Ref.~\cite{Shimizu:2019ptd}
only considers the longest-range part of the heavy
antimeson-baryon potential, i.e. OPE.
More recently, Ref.~\cite{Yamaguchi:2019seo} improves over the OPE calculation
of Ref.~\cite{Shimizu:2019ptd} by explicitly including
the $\bar{D} \Lambda_c$ and $\bar{D}^* \Lambda_c$
channels and a compact $c \bar{c} qqq$ core.
These improvements lead Ref.~\cite{Yamaguchi:2019seo} to predict the existence of
the full pentaquark septuplet and to determine that the quantum numbers of
the $P_c(4440)$ and $P_c(4457)$ are $J^P = \tfrac{3}{2}^-$
and $\tfrac{1}{2}^-$, i.e. scenario B.
But there are two important differences between Ref.~\cite{Yamaguchi:2019seo} and
the calculations in the present manuscript: (i) Ref.~\cite{Yamaguchi:2019seo}
takes $g_2 \sim 1.5$ (notice that they use the normalization of
Yan~\cite{Yan:1992gz} for the axial coupling,
where $g_{2} = \frac{3}{2} g_{1,\rm Yan}$),
(ii) the treatment of the short-range piece of
the interaction is phenomenological and is modeled with a compact
$c \bar{c} qqq$ core, which in turn leads to a short-range potential.

\section{Summary}
\label{sec:summary}

In this manuscript we have described the impact that pion exchanges
have in the description of the hidden-charm pentaquarks,
provided they are indeed molecular.
Pion exchanges are an important factor in the ordering of the pentaquark
spectrum, a factor that might determine which quantum numbers
are more/less bound.

If we try to describe consistently the LHCb pentaquark trio
with a pionful EFT, the preliminary conclusion is that
the $P_c(4440)$ and the $P_c(4457)$ are the
$J^P = \tfrac{3}{2}^-$ and $\tfrac{1}{2}^-$
$\bar{D}^* \Sigma_c$ molecular pentaquarks,
respectively.
This conclusion agrees with the previous work of
Karliner and Rosner~\cite{Karliner:2015ina},
which is not surprising once we take into account that
this is a consequence of OPE being attractive (repulsive)
in the  $\tfrac{3}{2}^-$ ($\tfrac{1}{2}^-$) channel.
But this identification is only marginally preferred
over the opposite one: the different uncertainties
within the pionful EFT description we use make
it impossible to reach a definite conclusion.
This is further compounded with the uncertainties in the location of
the $P_c(4312)$, $m = 4311.9 \pm 0.7 ^{+6.8}_{-0.7}$,
where the systematic uncertainty (i.e. the $^{+6.8}_{-0.7}$ error)
leans in the direction which results in a less bound molecular pentaquark.
The recent amplitude analysis of Ref.~\cite{Fernandez-Ramirez:2019koa},
which claims that the $P_c(4312)$ could be a virtual state,
cements this idea further.
If this is the case, the preferences of both scenarios
could likely change.

Besides the quantum numbers of the molecular pentaquarks, pion exchanges
lead to the prediction of a total of seven hidden-charm molecular
pentaquarks in the isodoublet $I=\tfrac{1}{2}$ sector.
This confirms the previous conclusions obtained
in a pionless EFT~\cite{Liu:2019tjn}, a more sophisticated pionless EFT
including coupled channels~\cite{Sakai:2019qph}, the hidden gauge model
(as constrained by HQSS)~\cite{Xiao:2019aya} and a recent
phenomenological pionful calculation~\cite{Yamaguchi:2019seo}.
In turn this points toward the idea that the existence of the HQSS multiplet
is more a consequence of HQSS than of the explicit dynamics leading to binding.
In particular the most important factor determining the details of
the binding energy is the quantum numbers of
the $P_c(4440)$ and $P_c(4457)$.

\begin{table*}[t]
\begin{tabular}{|c|c|c|c|c|}
\hline\hline
Molecule & Partial Waves & $J^P$ & $\vec{a}_1 \cdot \vec{a}_2 $ &
$S_{12} = 3\,\vec{a}_1 \cdot \hat{r} \, \vec{a}_2 \cdot \hat{r} -
\vec{a}_1 \cdot \vec{a}_2$ 
\\ \hline
$\bar{D} \Sigma_c$ & $^2S_{{1}/{2}}$ & $\frac{1}{2}^-$ & 0 & 0 \\ \hline
$\bar{D} \Sigma_c^*$ & $^4S_{{3}/{2}}$-$^4D_{{3}/{2}}$ &
$\frac{3}{2}^-$
& $\left(\begin{matrix}
0 & 0 \\
0 & 0 \\
\end{matrix}\right)$
& $\left(\begin{matrix}
0 & 0 \\
0 & 0 \\
\end{matrix}\right)$
\\ \hline
$\bar{D}^* \Sigma_c$ & $^2S_{{1}/{2}}$-$^4D_{{1}/{2}}$  & $\frac{1}{2}^-$ & 
$\left(\begin{matrix}
-2 & 0 \\
0 & 1 \\
\end{matrix}\right)$ & $\left(\begin{matrix}
0 & \sqrt{2} \\
\sqrt{2} & -2 \\
\end{matrix}\right)$ 
\\ \hline
$\bar{D}^* \Sigma_c$ & $^2D_{{3}/{2}}$-$^4S_{{1}/{2}}$-$^4D_{{1}/{2}}$
& $\frac{3}{2}^-$ & 
$\left(\begin{matrix}
-2 & 0 & 0 \\
0 & 1 & 0\\
0 & 0 & 1
\end{matrix}\right)$& $\left(\begin{matrix}
0 & -1 & 1 \\
-1 & 0 & 2 \\
1 & 2 & 0 
\end{matrix}\right)$ 
\\ \hline
$\bar{D}^* \Sigma_c^*$ & $^2S_{{1}/{2}}$-$^4D_{{1}/{2}}$-$^6D_{{1}/{2}}$
& $\frac{1}{2}^-$ & 
$\left(\begin{matrix}
-\frac{5}{2} & 0 & 0 \\
 0 & -1 & 0 \\
 0 & 0 & \frac{3}{2}
\end{matrix}\right)$ &
$\left(\begin{matrix}
 0 & -\frac{7}{2 \sqrt{5}} & -\frac{3}{\sqrt{5}} \\
 -\frac{7}{2 \sqrt{5}} & -\frac{8}{5} & -\frac{3}{10} \\
 -\frac{3}{\sqrt{5}} & -\frac{3}{10} & -\frac{12}{5}
\end{matrix}\right)$ 
\\ \hline
$\bar{D}^* \Sigma_c^*$ &
$^2D_{{3}/{2}}$-$^4S_{{3}/{2}}$-$^4D_{{3}/{2}}$-$^6D_{{3}/{2}}$-$^6G_{{3}/{2}}$ &
$\frac{3}{2}^-$ & 
$\left(\begin{smallmatrix}
-\frac{5}{2} & 0 & 0 & 0 & 0 \\
0 & -1 & 0 & 0 & 0 \\
0 & 0 & -1 & 0 & 0 \\
0 & 0 & 0 & \frac{3}{2} & 0\\
0 & 0 & 0 & 0 & \frac{3}{2} \\
\end{smallmatrix}\right)$
&
$\left(\begin{smallmatrix}
  0 & \frac{7}{2\sqrt{10}} & -\frac{7}{2\sqrt{10}} & \frac{3}{\sqrt{35}} &
  -3\,\sqrt{\frac{6}{35}} \\
  \frac{7}{2\sqrt{10}} & 0 & \frac{8}{5} & -\frac{3}{10}\sqrt{\frac{7}{2}} &
  0 \\
  -\frac{7}{2\sqrt{10}} & \frac{8}{5} & 0 & -\frac{3}{2 \sqrt{14}} &
  -\frac{3}{5}\,\sqrt{\frac{3}{7}} \\
  \frac{3}{\sqrt{35}} & -\frac{3}{10}\,\sqrt{\frac{7}{2}}
  & -\frac{3}{2\sqrt{14}} & -\frac{6}{7} & \frac{9 \sqrt{6}}{35} \\
  -3\,\sqrt{\frac{6}{35}} & 0 & -\frac{3}{5}\,\sqrt{\frac{3}{7}} &
  \frac{9 \sqrt{6}}{35} & -\frac{15}{7} \\
\end{smallmatrix}\right)$
\\ \hline
$\bar{D}^* \Sigma_c^*$ &
$^2D_{{5}/{2}}$-$^4D_{{5}/{2}}$-$^4G_{{5}/{2}}$-$^6S_{{5}/{2}}$-$^6D_{{5}/{2}}$-$^6G_{{5}/{2}}$
& $\frac{5}{2}^-$ &
$\left(\begin{smallmatrix}
-\frac{5}{2} & 0 & 0 & 0 & 0 & 0 \\
0 & -1 & 0 & 0 & 0 & 0\\
0 & 0 & -1 & 0 & 0 & 0\\
0 & 0 & 0 & \frac{3}{2} & 0 & 0\\
0 & 0 & 0 & 0 & \frac{3}{2} & 0\\
0 & 0 & 0 & 0 & 0 & \frac{3}{2} \\
\end{smallmatrix}\right)$
&
$\left(\begin{smallmatrix}
  0 & \frac{1}{2}\sqrt{\frac{7}{5}} & -\sqrt{\frac{21}{10}} &
  -\sqrt{\frac{3}{5}} & 2\sqrt{\frac{6}{35}} & -3\,\sqrt{\frac{2}{35}} \\
  \frac{1}{2}\sqrt{\frac{7}{5}} & \frac{8}{7} & \frac{16\,\sqrt{6}}{35}
  & \frac{\sqrt{21}}{10} & -\frac{1}{7}\sqrt{\frac{3}{2}} &
  - \frac{12 \sqrt{2}}{35} \\
  -\sqrt{\frac{21}{10}} & \frac{16\sqrt{6}}{35} & -\frac{8}{7} &
  0 & \frac{9}{70} & -\frac{3 \sqrt{3}}{14} \\
  -\sqrt{\frac{3}{5}} & \frac{\sqrt{21}}{10} & 0 & 0 &
  \frac{2 \sqrt{14}}{5} & 0\\
  2\sqrt{\frac{6}{35}} & -\frac{1}{7}\sqrt{\frac{3}{2}} & \frac{9}{70} &
  \frac{3 \sqrt{14}}{5} & \frac{6}{7} & \frac{27 \sqrt{3}}{35} \\
  -3\sqrt{\frac{2}{35}} & -\frac{12\sqrt{2}}{35} & -\frac{3\sqrt{3}}{14} &
  0 & \frac{27\sqrt{3}}{35} & -\frac{6}{7} \\
\end{smallmatrix}\right)$
\\ \hline
\hline\hline
\end{tabular}
\centering \caption{Matrix elements of the spin-spin and
  tensor operator for the partial waves we are considering
  in this work.} \label{tab:operators}
\end{table*}

\section*{Acknowledgments}
This work is partly supported by the National Natural Science Foundation of
China under Grants No. 11735003, the fundamental Research Funds
for the Central Universities, and the Thousand Talents Plan
for Young Professionals.


%

\end{document}